\documentclass[aps,prl,twocolumn,amsmath,amssymb]{revtex4-2}

\usepackage{lipsum}
\usepackage{graphicx}
\usepackage{dcolumn}
\usepackage{bm}
\usepackage{hyperref}
\usepackage{ulem}
\usepackage{textcomp}
\usepackage[version=4]{mhchem}
\hypersetup{hypertex=true,colorlinks=true,citecolor=blue,urlcolor=blue,linkcolor=blue,anchorcolor=blue}
\def\I{\uppercase\expandafter{\romannumeral 1}}
\def\II{\uppercase\expandafter{\romannumeral 2}}
\def\III{{\uppercase\expandafter{\romannumeral 3}}}
\def\IV{{\uppercase\expandafter{\romannumeral 4}}}
\def\V{{\uppercase\expandafter{\romannumeral 5}}}
\def\VI{{\uppercase\expandafter{\romannumeral 6}}}
\def\VII{{\uppercase\expandafter{\romannumeral 7}}}
\def\i{\lowercase\expandafter{\romannumeral 1}}
\def\ii{\lowercase\expandafter{\romannumeral 2}}
\def\iii{{\lowercase\expandafter{\romannumeral 3}}}
\def\iv{{\lowercase\expandafter{\romannumeral 4}}}
\def\v{{\lowercase\expandafter{\romannumeral 5}}}
\def\vi{{\lowercase\expandafter{\romannumeral 6}}}
\def\vii{{\lowercase\expandafter{\romannumeral 7}}}

\def\angstrom{\mbox{\normalfont\AA}}
\begin{document}

\title{Phonons in magic-angle twisted bilayer graphene}

\author{Xiaoqian Liu}
\affiliation{School of Physical Science and Technology, ShanghaiTech University, Shanghai 200031, China}

\author{Ran Peng}
\affiliation{School of Physical Science and Technology, ShanghaiTech University, Shanghai 200031, China}

\author{Zhaoru Sun}
\email[]{sunzhr@shanghaitech.edu.cn}
\affiliation{School of Physical Science and Technology, ShanghaiTech University, Shanghai 200031, China}

\author{Jianpeng Liu}
\email[]{liujp@shanghaitech.edu.cn}
\affiliation{School of Physical Science and Technology, ShanghaiTech University, Shanghai 200031, China}



\begin{abstract}
 Magic-angle twisted bilayer graphene (TBG) has attracted significant interest recently due to the discoveries of diverse correlated and topological states in this system. 
 Despite the extensive research on the electron-electron interaction effects and topological properties of the electrons, the phonons of magic-angle TBG are relatively less explored. In this work, we study the phonon properties in magic-angle TBG based on \textit{ab} \textit{initio} deep potential molecular dynamics. We have calculated phonon band structures and density of states at the magic angle, and have systematically analyzed the phonon eigenmodes at high-symmetry points in the moir\'e Brillouin zone. In particular, at the moir\'e $\Gamma$ point, we have discovered a number of soft modes which can exhibit  dipolar-like, stripe-like, and octupolar-like vibrational patterns within the moir\'e supercell, as well as some  ``vortical" modes with nonzero curl in real space. At the moir\'e $K$/$K'$ points, there are time-reversal breaking chiral phonon modes with nonzero local phonon polarizations. We have further studied the phonon effects on the electronic structures by freezing certain soft phonon modes. We find that if a soft ``stripe" phonon mode at moir\'e $\Gamma$ point is assumed to be frozen, the system would exhibit a charge order which naturally explains the recent observations from scanning tunnelling microscopy. Moreover, there are also low-frequency $C_{2z}$-breaking modes at moir\'e $\Gamma$ point, which would gap out the Dirac points at the charge neutrality point once these modes get frozen. This provides a new perspective to the origin of correlated insulator state at the charge neutrality point.

\end{abstract}

\maketitle

Twisted bilayer graphene (TBG) system around the magic angle is an ideal platform to realize various intriguing quantum phases \cite{balents-review-tbg,andrei-review-tbg} such as the correlated insulators \cite{cao-nature18-mott,efetov-nature19,tbg-stm-pasupathy19,tbg-stm-andrei19,tbg-stm-yazdani19, tbg-stm-caltech19, young-tbg-science19,efetov-nature20,young-tbg-np20,li-tbg-science21,yacoby-fqhe-tbg-arxiv21,efetov-tbg-flux-arxiv21}, quantum anomalous Hall states \cite{young-tbg-science19, sharpe-science-19, efetov-arxiv20,yazdani-tbg-chern-arxiv20,andrei-tbg-chern-nm21,efetov-tbg-chern-arxiv20,pablo-tbg-chern-arxiv21,yacoby-fqhe-tbg-arxiv21,shen-tbg-chern-cpl21,choi-tbg-chern-nature21,efetov-tbg-prl21,jpliu-nrp21},  unconventional superconductivity\cite{cao-nature18-supercond,dean-tbg-science19,marc-tbg-19, efetov-nature19,efetov-nature20,young-tbg-np20,li-tbg-science21,cao-tbg-nematic-science21,efetov-tbg-prl21}, and linear-in-temperature resistivity.  Around magic angle 1.05$^{\circ}$\cite{macdonald-pnas11}, there are two topologically nontrivial flat bands contributed by each valley and spin degrees of freedom \cite{song-tbg-prl19, yang-tbg-prx19,po-tbg-prb19, origin-magic-angle-prl19, jpliu-prb19}. A lot of the unusual phenomena, including correlated insulators and quantum anomalous Hall effects,  can be attributed to the presence of such topologically nontrivial flat bands in the electronic degrees of freedom. The electron-electron ($e\rm{-}e$) Coulomb interactions dominates over the  kinetic energy near magic angle, and the interplay between the strong Coulomb correlations and the nontrivial topology of the flat bands give rise to  diverse correlated insulator states and topological states, which have been extensively studied from the theoretical point of view over the past few years \cite{po-prx18,koshino-prx18,kang-prx18,zou-tbg-prb18,yuan-tbg-prb18,kang-tbg-prl19,Uchoa-ferroMott-prl,xie-tbg-2018, zaletel-tbg-2019,zhang-senthil-tbg19,zaletel-tbg-hf-prx20,jpliu-tbghf-prb21,zhang-tbghf-arxiv20,hejazi-tbg-hf,kang-tbg-dmrg-prb20,wu-tbg-collective-prl20,kang-tbg-topomott,yang-tbg-arxiv20,meng-tbg-arxiv20,he-nc20,lin-tbg-prl20,eslam-soft-mode-arxiv20,Bernevig-tbg3-arxiv20,Lian-tbg4-arxiv20,bernevig-tbg-5-prb21,regnault-tbg-ed,zaletel-dmrg-prb20,huang-orbit-prl21,macdonald-tbg-ed-arxiv21,meng-tbg-qmc-cpl21,lee-tbg-qmc-arxiv21,bultinck-tbg-strain-prl21,xie-tbg-prl21,balents-tbg-prb21,wagner-iks-arxiv21,kwan-iks-arxiv21,zhang-nonlinear-arxiv21,bernevig-spectroscopy-tbg-arxiv21,bernevig-tbg-flux-arxiv21,song-tbg-heavy-arxiv21,shi-tbg-exciton-arxiv21}. 

However, some other phenomena observed in magic-angle TBG, such as unconventional superconductivity \cite{cao-nature18-supercond,dean-tbg-science19,marc-tbg-19, efetov-nature19,efetov-nature20,young-tbg-np20,li-tbg-science21,cao-tbg-nematic-science21,efetov-tbg-prl21} and linear-in-temperature resistivity \cite{young-linear-np19,cao-strange-prl20}, are relatively less understood. One of the perspectives is that these unusual transport phenomena may result from electron-phonon couplings \cite{wu-prl18,lian-prl18,wu-linear-prb19,sharma-tbg-phonon-nc21}, and are thus intimately related to phonon vibrational properties. In particular, recent experiments show that the correlated insulator states that are driven by $e\rm{-}e$ interactions tend to compete with superconductivity, and the latter is getting suppressed when the $e\rm{-}e$ interaction strength increases \cite{li-tbg-science21,young-tbg-np20}. This seems to imply that electron-phonon interactions are crucial in order to understand the origin of superconductivity in this system. On the other hand, despite a few pioneering works \cite{balandin-prb13,choi-prb18,wu-prl18,lian-prl18,tbg-raman-nc18,wu-linear-prb19,angeli-tbg-prx19,koshino-tbg-prb19,koshino-prb20,sharma-tbg-phonon-nc21,choi-tbg-prl21},  the phononic properties and electron-phonon couplings in magic-angle TBG are much less explored compared to the comprehensive research on the electronic degrees of freedom.  

In such a context, in this work we  study the phonon properties of magic-angle TBG based on \textit{ab} \textit{initio} deep potential molecular dynamics (DPMD) method.
To be specific, a machine-learning-based reactive potential method \cite{wangh-dpkit-comphy18,zhanglf-dpgen-comphy20} is adopted, 
which allows for an accurate multi-body  description for the interatomic potentials of large-scale systems such as magic-angle TBG (with $\sim 11000$ atoms in each moir\'e primitive cell). The accuracy of the calculated total energies and forces based on such DPMD method is comparable to that from first principles calculations based on density functional theory (DFT).
Using this method, we have calculated phonon band structures and phonon density of states at the magic angle, and have systematically analyzed the phonon eigenmodes at high-symmetry points in the moir\'e Brillouin zone. In particular, at the moir\'e $\Gamma$ point, we have discovered a number of soft phonon modes with frequencies $\sim 0.05\textrm{-}0.1\,$THz, which exhibit various intriguing vibrational patterns on the moir\'e length scale. 
At the moir\'e $K$/$K'$ points , there are  time-reversal breaking chiral phonon modes with nonzero local phonon polarizations, which may be coupled with the orbital motions of electrons and boost the formation of orbital magnetic ground states. We have further studied the phonon effects on the electronic structures by freezing certain soft phonon modes. We find that if a soft stripe-type phonon mode were frozen,  the system would exhibit a charge order which naturally explains the recent observations from scanning tunnelling microscopy (STM) \cite{tbg-stm-andrei19}. Moreover,  if some of the soft $C_{2z}$-breaking modes at $\Gamma$ point were frozen, the electronic band structure becomes gapped at the charge neutrality point (CNP), which may provide a new perspective to the origin of correlated insulator state at CNP observed in experiments.

\begin{figure}[htbp]
	\centering\label{Fig1}
	\includegraphics[width=3.5in]{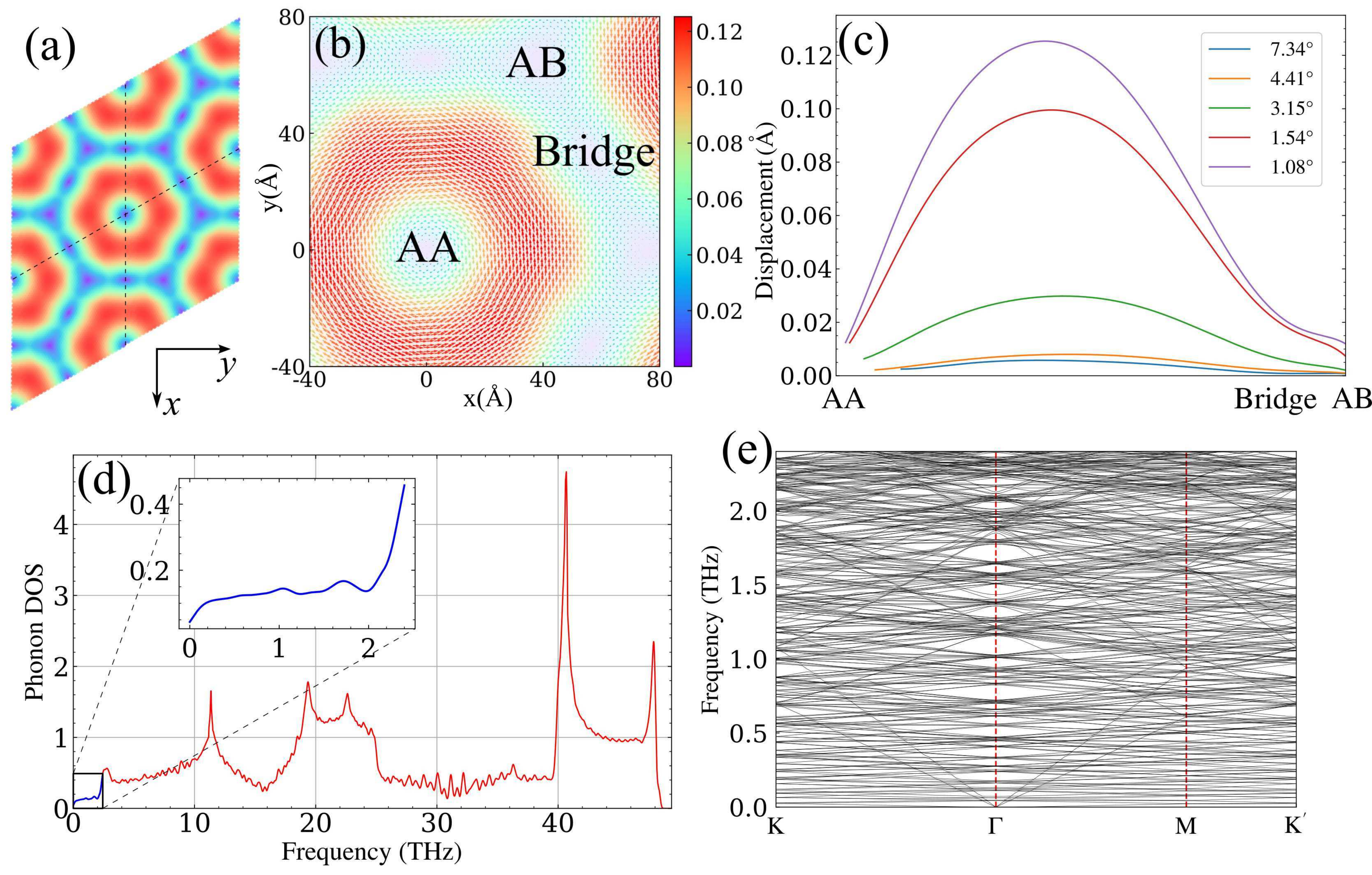}
	\caption{Structures of TBG after geometric optimization. (a) An overview of the relaxed structure, and (b) a zoomed in view of the relaxed structure, with the color coding and arrows denoting the amplitudes and directions of the in-plane atomic displacement vectors. (c) Amplitudes of atomic displacement vectors plotted along a line connecting $AA$-bridge-$AB$ points (marked in (b)) , where the twist angle is at 7.43$^{\circ}$, 4.41$^{\circ}$, 3.15$^{\circ}$, 1.54$^{\circ}$, and 1.08$^{\circ}$. (d) The phonon density of states (DOS) of magic-angle TBG, where the inset shows the low-frequency DOS from 0\,THz to 2.4\,THz. (e) Phonon dispersions of magic-angle TBG from 0 to 2.4\,THz.}\label{fig1}
\end{figure}

We first apply the DPMD method to the problem of structural relaxation of magic-angle TBG.  In order to obtain a reliable interatomic potential, we first performed first principles calculations for TBG based on DFT as implemented by the Vienna Ab initio Simulation Package (VASP)\cite{vasp} , with the twist angle $\theta$ varying from 21.79$^{\circ}$ to 4.41$^{\circ}$. In particular, for the moir\'e primitive cell at each twist angle ($21.79^{\circ}\leq\theta\leq\!4.41^{\circ}$), we have performed \textit{ab} \textit{initio} MD simulations for at least 100\,fs to obtain plenty enough training data.  Then the reliable interatomic potential based on neural networks can be trained from DFT energies and forces, based on which we further study the structural and phononic properties of TBG at smaller twist angles (including the magic angle $\theta\!\approx\!1.08^{\circ}$) with larger moir\'e primitive cells with the accuracy comparable to direct DFT simulations. 
The details of this method are presented in Supplementary Information \cite{supp_info}. 
The results of the structural relaxations are presented in Fig.~\ref{fig1}. To be specific, in Fig.~\ref{fig1}(a) we show an overview of the relaxed structure of magic-angle TBG, where the color coding indicates the amplitudes of the atomic displacements deviating from the ideal moir\'e lattice positions obtained by a simple twist of the two graphene monolayers centered at the $AA$ point. We see that the atomic displacements have largest amplitudes in a the region between the $AA$ and $AB/BA$ points, forming a ring encircling the $AA$ region. Further analysis reveal that these atomic displacement vectors of the relaxed structure are ``winding" around the $AA$ point, generating a vortex-like in-plane vector field, as clearly shown by the arrows (representing the directions of the in-plane displacements) in Fig.~\ref{fig1}(b). Moreover, as $AB/BA$ stacked bilayer graphene is energetically more stable than the $AA$ stacked one, carbon atoms near the $AA$ point tend to misalign with each other, resulting in a squeeze of the $AA$ region.  In addition to the in-plane displacements, our calculations indicate that there are also out-of-plane corrugations, i.e., the periodic modulations of interlayer distance in different regions of the moir\'e pattern, with the distance at the $AA$ ($AB/BA$) point being 3.62\,\angstrom (3.36\,\angstrom), which is consistent with previous results \cite{tbg-corrugation-prb14,angeli-d6-prb18,koshino-prb17,tbg-relax-prr19}. 

We have further calculated the atomic displacements of TBG with different twist angles. In Fig.~\ref{fig1}(c) we show the amplitudes of the atomic displacements in the relaxed TBG structures at different twist angles along a line  connecting the $AA$-bridge-$AB$ points. At large twist angles, $\theta\gtrapprox 4^{\circ}$, the atomic displacements in the relaxed structure are very weak $\sim0.005\,$\angstrom; but for smaller angles, the displacement amplitudes are greatly enhanced. Around the magic angle, the maximal displacement reaches $0.12\,$\angstrom.  Such significant structural relaxations are crucial in determining the low-energy electronic structures of magic-angle TBG.   


\begin{figure*}[htbp]
	\centering
	\includegraphics[width=5in]{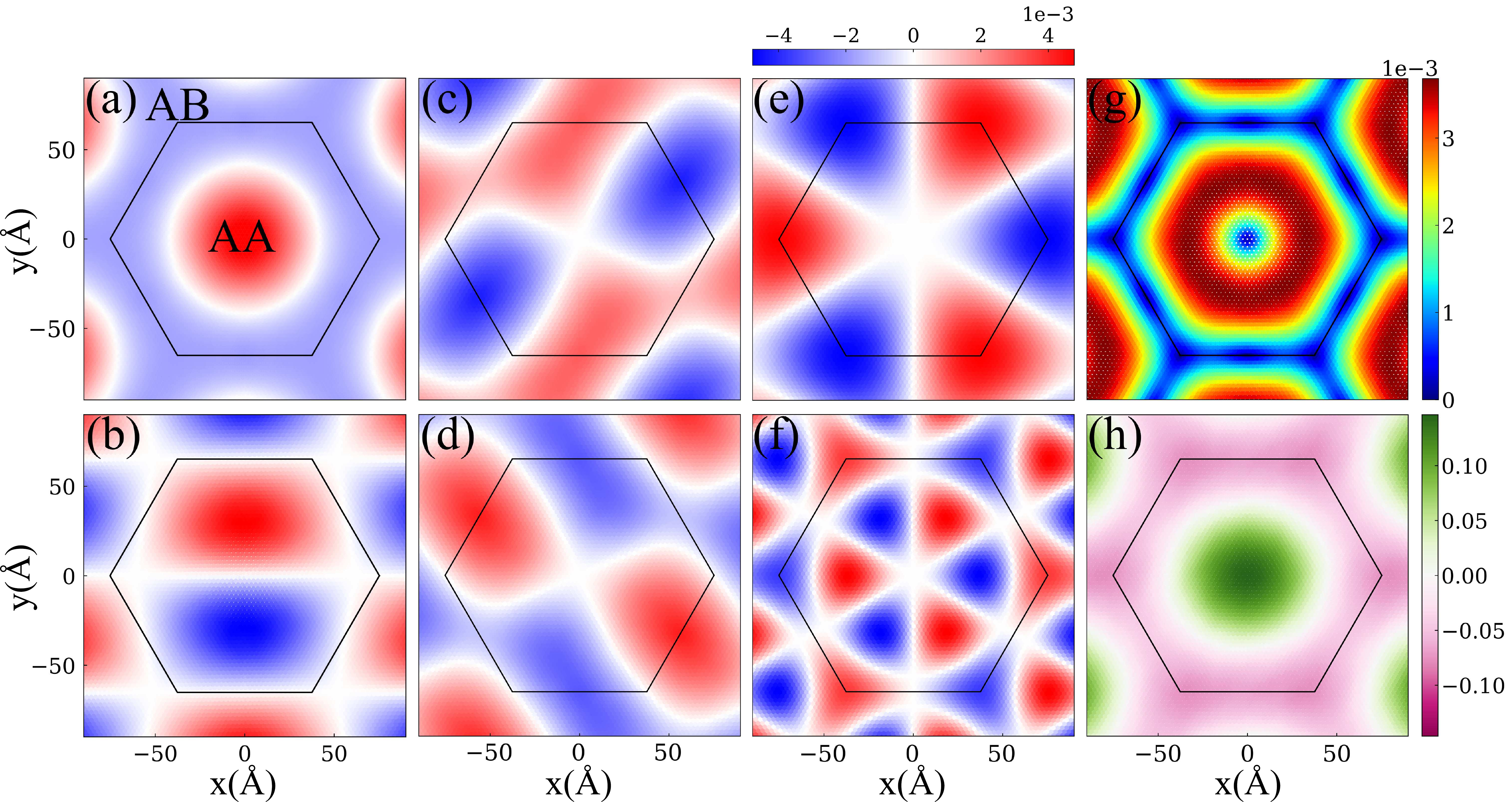}
	\caption{Low frequency optical phonon modes at $\Gamma$. (a)-(f) The out-of-plane (normalized) vibrational amplitudes for the soft phonon modes with frequencies at $0.043\,$, $0.046\,$, $0.063\,$,  $0.063\,$, $0.073\,$, and $0.163\,$THz, respectively, where the black hexagon marks the moir\'e primitive cell. (g) The inplane vortex pattern of vortical mode with frequency $1.218\,$THz. (h) The curl field of the vortical mode.}  \label{fig2}
\end{figure*}


Based on the fully relaxed structure, we further study the phonon properties of magic-angle TBG. In Fig.~\ref{fig2}(a) we show the calculated phonon density of states for magic-angle TBG, which exhibit a few peaks around $11.35$, $19.36$, $40.68$, and $47.88\,$THz, consistent with previous reports \cite{choi-prb18}. Despite slight shift of DOS peaks, the phonon DOS of magic-angle TBG at the first glance are quite similar to those of bilayer graphene \cite{choi-prb18,balandin-prb13}. However, if one closely looks at the low-frequency phonon modes with long-wavelength vibrational patterns, it would be completely different. In magic-angle TBG, one would obtain numerous low-frequency moir\'e phonon modes which cannot be simply interpreted as folding atomic phonon modes of bilayer graphene into the moir\'e Brillouin zone. In this work, we specifically focus at the low-frequency, long-wavelength moir\'e phonon properties with frequency $\sim\!0\textrm{-}2.4\,$THz, as marked by the blue lines in Fig.~\ref{fig1}(d) (the inset shows zoom-in phonon DOS in this low-frequency regime). In Fig.~\ref{fig1}(e) we further show the phonon dispersions 
of magic-angle TBG with the frequency ranging from 0 to 2.4\,THz. There are already hundreds of phonon modes within such a small frequency regime, and some of the optical phonon modes are extremely soft $\sim\!0.01\textrm{-}0.1\,$Thz, implying that the system is likely to undergo structural transitions, which may significantly change the electronic structure. In what follows we will comprehensively analyze the low-frequency optical phonon modes of magic-angle TBG at the high-symmetry points in the moir\'e Brillouin zone. 

We first focus at the phonon modes at $\Gamma$.  In Fig.~\ref{fig2} we show the vibrational patterns of several prototypical low-frequency optical phonon modes at $\Gamma$. To be specific, in Fig.~\ref{fig2}(a) we show the out-of-plane vibrational pattern of the lowest optical phonon at $\Gamma$ point with frequency $0.043\,$THz, which exhibit maximal amplitudes in the $AA$ region. In Fig.~\ref{fig2}(b) we show the  out-of-plane vibrational amplitudes of the second optical phonon mode with frequency $0.046\,$THz, which exhibits  a dipolar vibration pattern that clearly breaks $C_{2z}$ and $C_{3z}$ symmetries. This mode is doubly degenerate with another similar dipolar mode forming a two dimensional representation of $C_{3z}$ operation. In Fig.~\ref{fig2}(c) we show the out-of-plane vibration amplitudes of the two degenerate  ``stripe-type" modes with frequency $0.063\,$THz. As will be discussed later, if these two modes were frozen, the system would generate a stripe charge order that is consistent with the STM observation reported in Ref.~\onlinecite{tbg-stm-andrei19}.  In Fig.~\ref{fig2}(e) and (f) we show two ``octupolar-type" out-of-plane vibrational modes with the frequencies being $0.073\,$THz and $0.163\,$THz respectively. These two modes break $C_{2z}$ symmetry, and may open up a gap at the CNP. In Fig.~\ref{fig2}(g) we show the in-plane vibration amplitudes of another optical phonon mode with frequency $\sim 1.218\,$THz. The atomic displacement vectors show an interesting vortex pattern winding around the $AA$ point, which exhibits non-vanishing curl as shown in Fig.~\ref{fig2}(h).

\begin{figure}[htbp]
	\centering
	\includegraphics[width=3.5in]{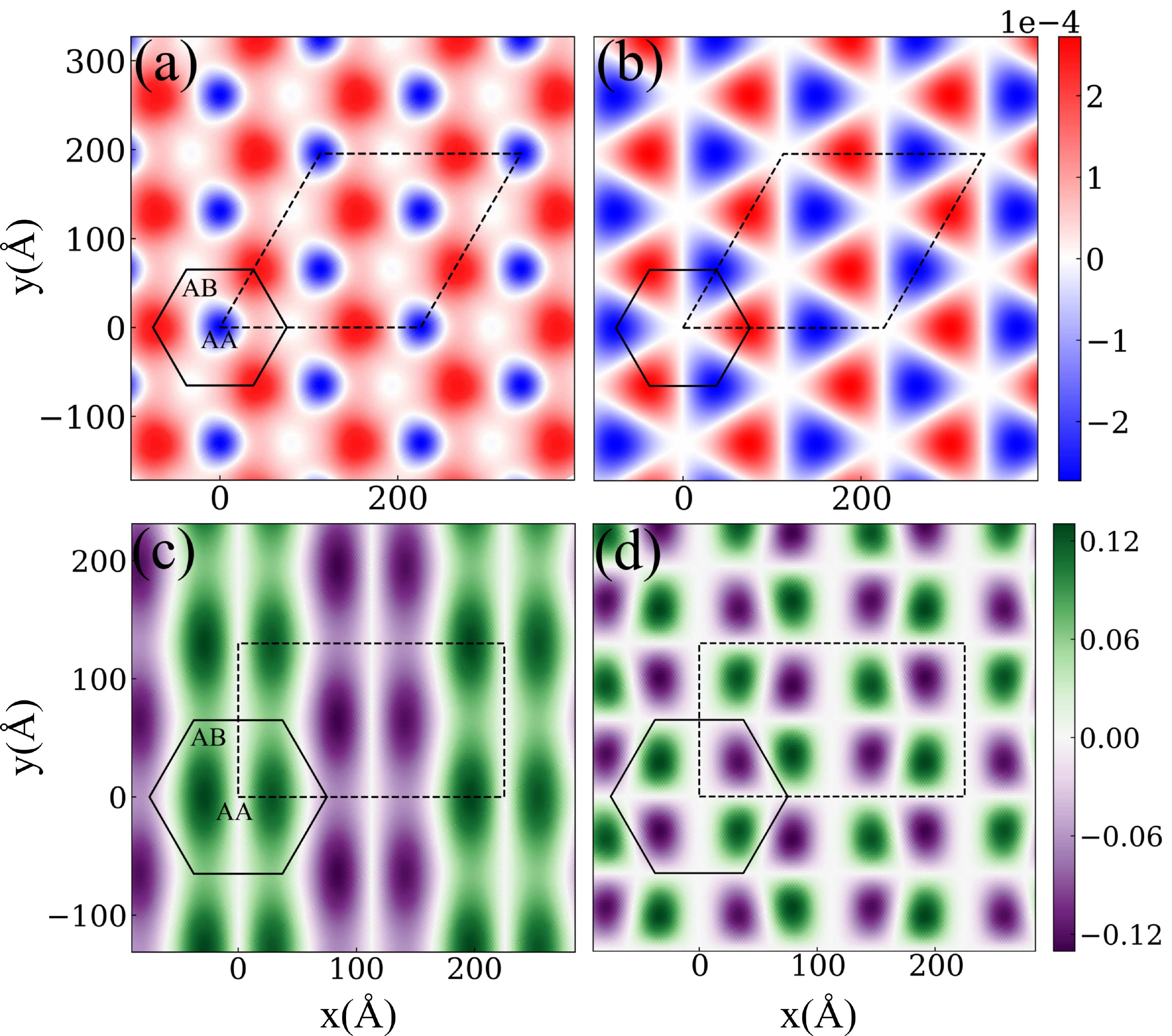}
	\caption{The phonon polarizations at  $K$ point and soft modes at $M$ point. (a)-(b): The local phonon polarizations of the optical modes at $K$ point with frequencies $1.214\,$THz and $1.074\,$THz respectively, where the primitive cell and the $\sqrt{3}\times\sqrt{3}$ moir\'e supercell are marked with solid line and dashed lines. (c)-(d): The out-of-plane vibrational amplitudes of the soft modes at $M$ point with frequencies at $1.214\,$THz and $1.074\,$THz respectively. The primitive cell and doubled moir\'e supercell are marked with black  and dashed lines.} \label{fig3}
\end{figure}

We continue to discuss the phonon properties at other high symmetry points in the moir\'e Brillouin zone. Interestingly, at moir\'e $K$ point we find that there are doubly degenerate chiral phonon modes of opposite chiralities, as well as some non-degenerate ``helical" phonon modes which have vanishing net chirality but  nontrivial local distribution of phonon polarizations. 
To be specific, the chirality of a phonon mode can be characterized by its polarization denoted by $\eta$, which is obtained by projecting a phonon eigenvector $\mathbf{u}_{n}(\mathbf{q})$ ($n$ the phonon mode index and $\mathbf{q}$ is the wavevector) onto the right-handed and left-handed basis vectors, then make the subtraction \cite{zhanglf-chiralphonon-prl15}:
$\eta=\sum_{i} \eta_i=\sum_i(\vert\langle R_i\vert \mathbf{u}_n(\mathbf{q})\rangle\vert^2-\vert\langle L_i\vert \mathbf{u}_n(\mathbf{q})\rangle\vert^2))$
%
where $\vert R_i\rangle\!=\!(1/\sqrt{2})\times[1,i]^{\textrm{T}}$ and $\vert L_i\rangle\!=\!(1/\sqrt{2})\times[1,-i]^{\textrm{T}}$ are the right-handed and left-handed basis vectors for the displacement of the $i$th atom (``T" stands for transpose conjugation). 
In Fig.~\ref{fig3}(a) we show the calculated local distribution of phonon polarization  of one of the doubly degenerate chiral modes at $K$ with frequency $1.214\,$THz, where the dashed line marks a $\sqrt{3}\times\sqrt{3}$ moir\'e supercell, and the solid black hexagon marks the moir\'e primitive cell. We see the phonon polarization reaches the maximal value in the $AA$ region, and the net polarization of each moir\'e supercell $\sim\!0.38$. In Fig.~\ref{fig3}(b) we show the local distribution a non-degenerate ``helical" phonon mode with frequency $\sim\!1.074\,$THz,  which has an octupolar-type distribution of local polarization with vanishing net chirality. It is worthwhile to note that although the atomic displacement vectors of a phonon mode at $K$ (or $K'$) point breaks the  primitive moir\'e translational symmetry, the phonon polarization still preserves the original translational symmetry as can be seen from the definition of phonon polarization. These intriguing time-reversal breaking phonon modes at $K/K'$ points may be coupled with the electronic degrees of freedom, and may help in the formation of orbital magnetic states breaking moir\'e translational symmetry.

In Fig.~\ref{fig3}(c) and (d) we show the real-space distributions of the out-of-plane vibrational amplitudes of the first and seventh optical phonon modes with extremely low frequencies $0.021\,$THz and $0.082\,$THz respectively. The out-of-plane atomic displacement vectors of the first phonon mode  at $M$ point form a stripe pattern with anti-phase modulation along a direction perpendicular to the lattice vector. Such an extremely soft phonon mode may be relevant with the unusual zero-Chern-number insulator states observed at filling factors 1 and 3 of the flat bands \cite{efetov-nature20}. Once this soft mode is frozen, the system has to break moir\'e translational symmetry in such a way to double the moir\'e primitive cell as indicated by the black dashed rectangle, and within the doubled moir\'e supercell there would be two holes (six holes) at filling factor 1 (3) for each doubled supercell, which has the chance to realize a zero-Chern-number insulator state without the necessity to completely lift the valley-spin degeneracy of the topologically nontrivial flat bands.  In Fig.~\ref{fig3}(d) we show the real-space distribution of the out-of-plane vibrational amplitudes of the third phonon mode at $M$ point, which exhibit an interesting quadrupolar pattern. 

\begin{figure}[htbp]
	\centering
	\includegraphics[width=3.5in]{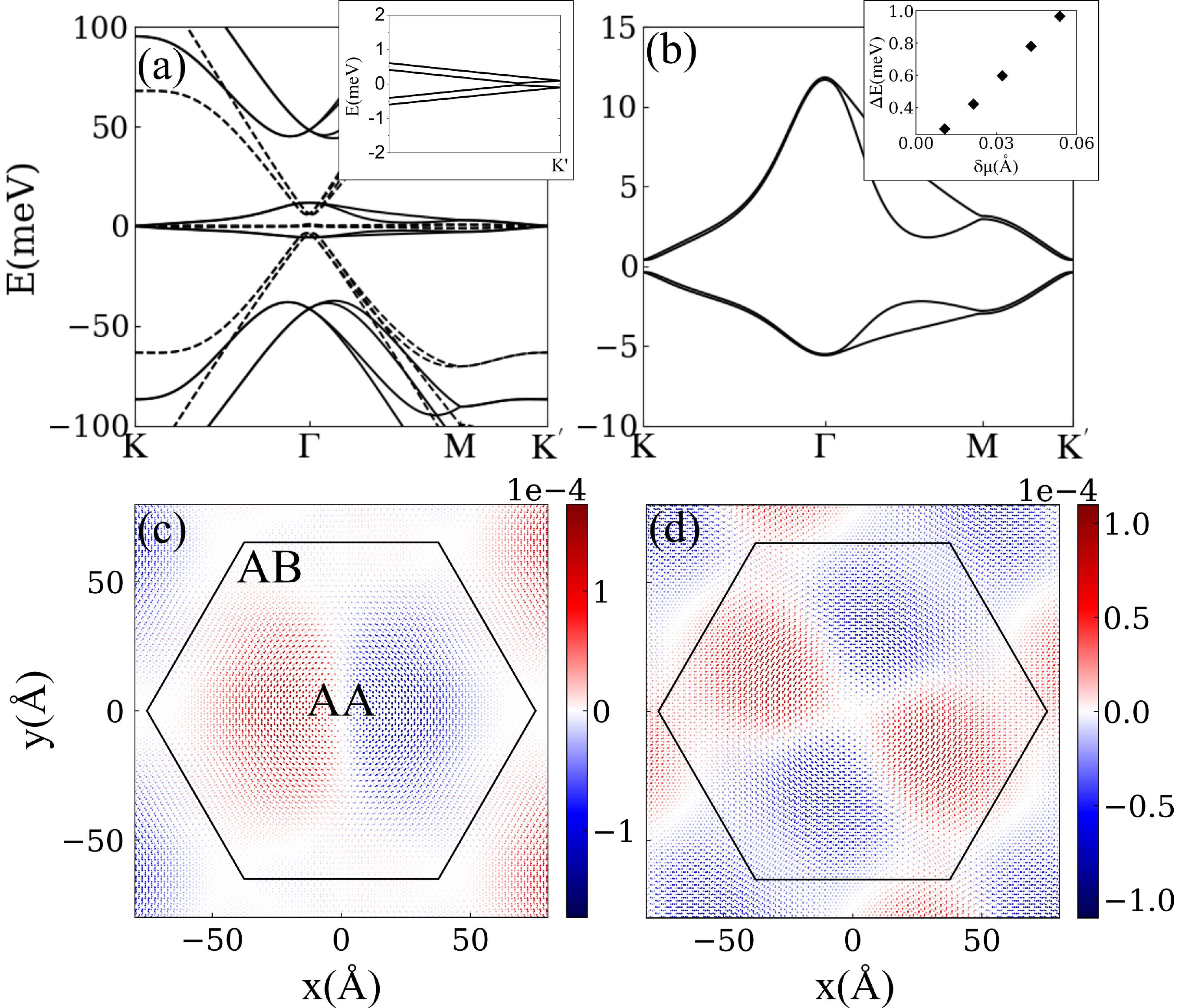}
	\caption{(a): Band structure of relaxed (solid lines) and unrelaxed (dashed lines) magic-angle TBG, where the inset indicates the gap opening at the Dirac point. (b): Flat bands of magic-angle TBG with the octupolar-type phonon modes shown in Fig.~2(f) ($0.163\,$THz) under frozen mode approximation, with the average displacement amplitudes by the mode of 0.0430\,\angstrom\. The inset shows the gap at moir\'e $K$ and $K^\prime$ points as a function of average displacement amplitudes. (c)-(d): The local charge distribution with the two soft modes shown in Fig.~1(b) (0.046\,THz) and Fig.~1(c) (0.063\,THz) being frozen.
	} \label{fig4}
\end{figure}

Before discussing the effects of the phonon modes, the structural relaxation effects on the electronic band structure of magic-angle TBG should be first understood. In Fig.~\ref{fig4}(a) we show the band structures of magic-angle TBG before and after relaxation as marked by the dashed and solid lines respectively. Clearly the lattice relaxation enhances both the gaps between remote bands and flat bands and the bandwidth of the flat bands. In the while the intervalley scattering is significantly enhanced, generating a splitting of approximately $0.19\,$meV between the two Dirac points (see inset of Fig.~\ref{fig4}(a)). 
We continue to study the electron-phonon coupling effects on the electronic degrees of freedom by calculating the electronic band structures under the ``frozen-phonon" approximation. 
In Fig.~\ref{fig4}(b) we show the band structure in a lattice structure with an octupolar-type phonon mode around $0.163\,$THz (Fig.~\ref{fig3}(f)) being frozen. A gap of $\sim 0.75\,$meV is created with an average displacement amplitude  $\sim0.05\,\angstrom$, as shown in the inset of Fig.~\ref{fig4}(b).  The coupling between the soft phonon modes and the electrons can also significantly change the local charge distributions, which may explain previous STM observations \cite{tbg-stm-andrei19}. In particular, we have calculated the local charge distribution contributed by the valence flat bands at each atomic site, with the ``stripe-type" phonon mode shown in Fig.~\ref{fig2}(c) ($\sim$0.063\,THz) being frozen, presented in Fig.~\ref{fig4}(d), after subtracting the charge distribution of the original relaxed structure without any phonon displacement. The difference of charge distribution forms a quasi-1D stripe pattern, which is in perfect agreement with previous STM observation \cite{tbg-stm-andrei19}. Similarly, we have also calculated the local charge distribution contributed by the valence flat bands with the ``dipolar" phonon mode ($\sim\!0.046\,$THz, shown in Fig.~\ref{fig2}(b)) being frozen, and the local charge distribution (also with the original charge distribution substracted) shows a diploar pattern as shown in Fig.~\ref{fig4}(c).  These results indicate that the phonon modes and electron-phonon couplings play a crucial role in the correlated states of magic-angle TBG. 

To summarize, in this work we have comprehensively studied the phononic properties of magic-angle TBG based on DPMD calculations. We have discovered a number of low-frequency  optical phonon modes at the high-symmetry points of moir\'e Brillouin zone, which exhibit various dipolar, stripe-like, octupolar, vortical, and chiral vibrational patterns on the moir\'e length scale. We have further studied the electron-phonon couplings by freezing certain types of soft phonon modes and study their effects on electronic band structures. We find that if a soft stripe-type phonon mode were frozen,  the system would exhibit a charge order which naturally explains the recent STM observation. Moreover,  the freezing of certain $C_{2z}$-broken phonon mode would open a gap at CNP, which may provides a new perspective to the origin of correlated insulator at CNP observed in experiment. Our work is a significant step forward in understanding the phononic and electronic properties of magic-angle TBG, and will provide useful guidelines for future experimental and theoretical studies.

\section*{Acknowledgement}
R. P. and J. P. L. acknowledge support from National Key R \& D program of China (grant no. 2020YFA0309601) and the National Science Foundation of China (grant no. 12174257). Computational resources are provided by the High-Performance Computing (HPC) Platform at ShanghaiTech University and Shanghai HPC center. We thank the useful discussion from Kexin Zhang and Shihao Zhang.

\bibliographystyle{apsrev4-1}
\bibliography{tmg}

\widetext
\clearpage

\begin{center}
\textbf{\large Supplementary Information of ``Phonons in magic-angle twisted bilayer graphene"}
\end{center}


\pagenumbering{roman}

\vspace{12pt}
\begin{center}
\textbf{\large \I\ Relaxed structure of twisted bilayer graphene}
\end{center}

\begin{figure}[htbp]
	\centering
	\includegraphics[width=3.5in]{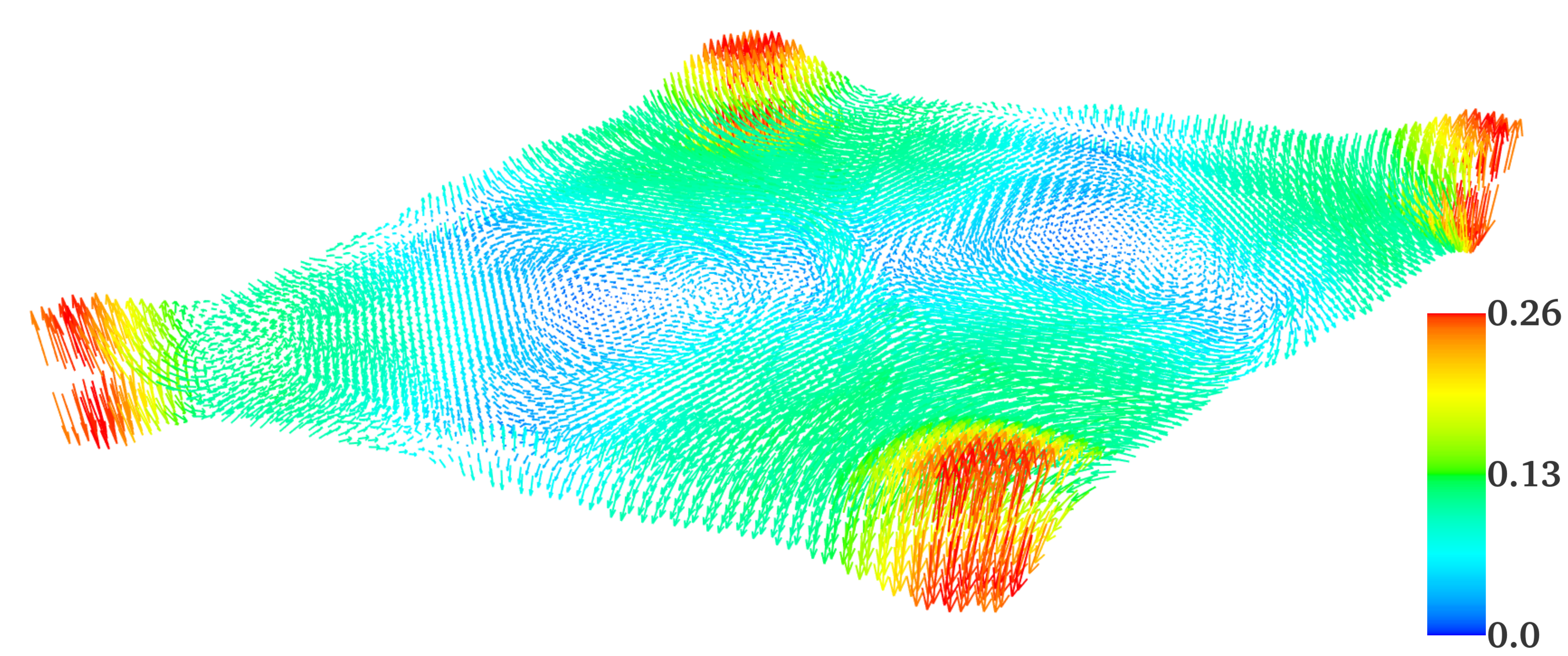}
	\caption{Atomic displacement vectors of magic-angle TBG after structural relaxation.}\label{fig_relax}
\end{figure}

\begin{figure}[htbp]
	\centering
	\includegraphics[width=2.5in]{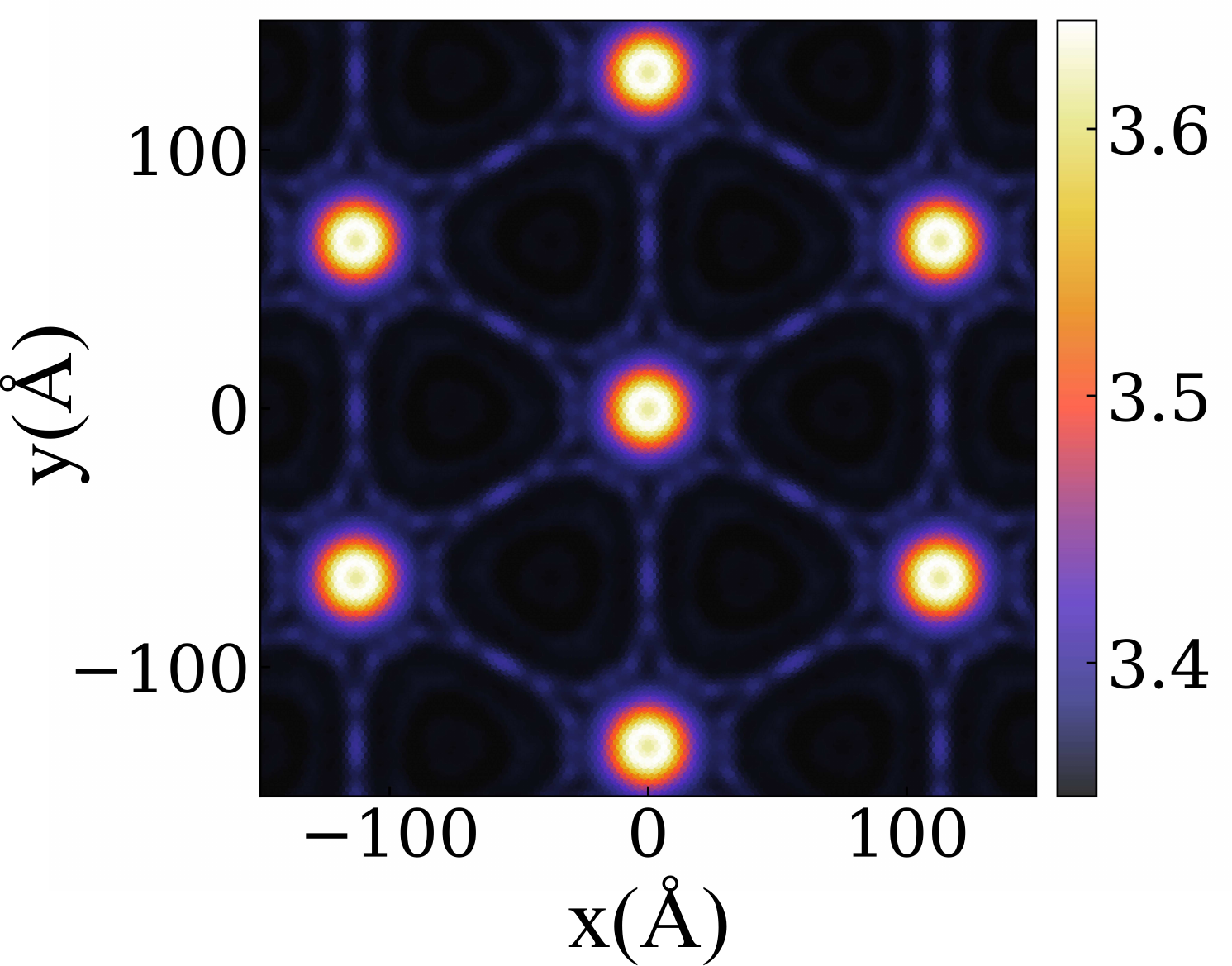}
	\caption{The interlayer distance as a function of real-space coordinates $(x,y)$ in the moir\'e superlattice.}\label{fig_outplane}
\end{figure}

The atomic displacement vectors in the relaxed lattice structure of magic-angle TBG have been presented in Fig.~\ref{fig_relax}, where the color coding and the vector fields indicate the amplitudes and directions of the atomic displacements. Inspection shows that after structural relaxation  the interlayer distance in the $AA$ region tends to be increased, leading to the out-of-plane corrugation. In particular, the interlayer distance at the $AA$ point is as large as 3.62\,\angstrom, while at the $AB/BA$ point it is 3.36\,\angstrom, consistent with previous report \cite{tbg-corrugation-prb14,giovanni-tbg-dft-prb19}. This is clearly shown in Fig.~\ref{fig_outplane}, where the color coding denotes the in-plane variation of the interlayer distance in the moir\'e supercell. 
Moreover, the $AA$ ($AB/BA$) region is contracted (expanded) after structural relaxation. The in-plane atomic displacement vectors wind around the $AA$ point, forming a vortex-like pattern. 

\vspace{12pt}
\begin{center}
\textbf{\large \II\ Deep potential molecular dynamics}
\end{center}

\begin{figure}[htbp]
	\centering
	\includegraphics[width=3.5in]{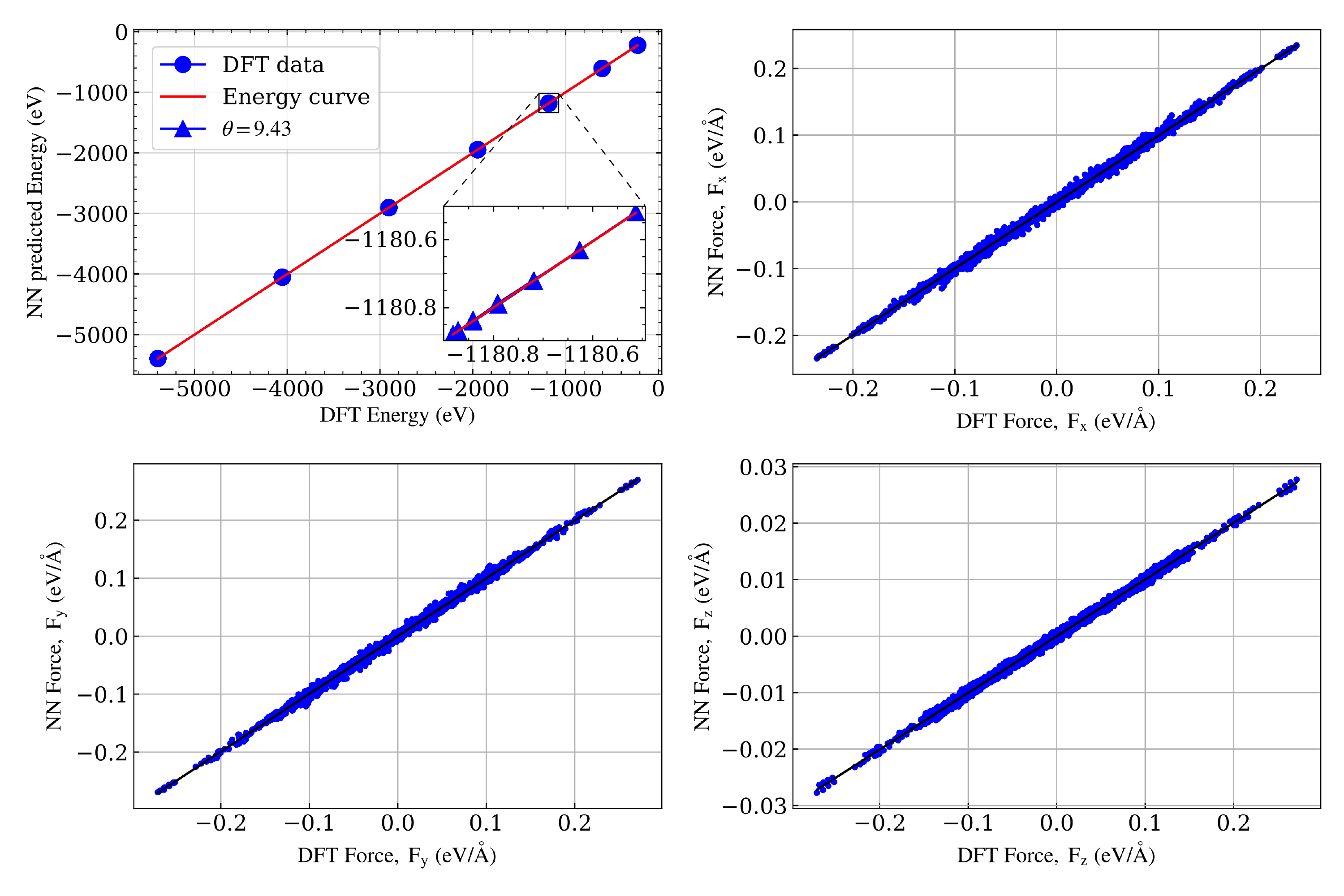}
	\caption{Validation of the neural network (NN) potential: (a) Comparison of total energies calculated from NN potential and DFT.deep-potential model prediction; (b-d) Comparison of forces calculated from NN potential and DFT in $x,y,z$ directions, respectively.} \label{fig_NN}
\end{figure}

Primitive moir\'e supercells with different twist angles (from 21.79\textdegree to 4.41\textdegree) have been constructed to get complex enough local atomic environments \cite{castro-neto-prb12}. The rotation center is chosen to be the center of the hexagon formed by the carbon atoms. With such a choice of rotation center, the $D_6$ symmetry of graphene is preserved for the TBG system.
In order to describe the tiny lattice distortions of TBG during structural relaxations, we constructed several structures with small perturbative atomic displacements from ideal moir\'e pattern as the initial configurations for ab initio molecular dynamics (AIMD). For each configurations, we performed AIMD calculation for at least 100\,fs to obtain plenty enough training data. The  the AIMD simulation have been carried out using the Vienna Ab initio Simulation Package (VASP) \cite{vasp} with vdW-DF2 exchange-correlation functional \cite{vdw-df2}. $D_6$ symmetry is enforced in the AIMD simulation. Under periodic conditions, a vacuum layer of 20\,\angstrom\ in the $z$ direction has been set up. Then the reliable interatomic potential based on neural networks can be trained from DFT energies and forces, based on which we further studied the structural and phononic properties of TBG at smaller twist angles (including the magic angle $\theta\!\approx\!1.08^{\circ}$) with larger moir\'e primitive cells with the accuracy comparable to DFT calculations. Otherwise it would be extremely demanding to  directly tackle with the structural properties of magic-angle TBG with DFT accuracy. 
The structural relaxation and force-constant matrix calculations were calculated by Large-scale Atomic/Molecular Massively Parallel Simulator (LAMMPS) \cite{lammps} with such NN potential\cite{wangh-dpkit-comphy18}. Finally, based on the force-constant matrix, phonons were calculated by phonopy \cite{phonopy} package with a $2\times2$ moir\'e supercell. 
The accuracy of our NN potential has been validated by comparing the total energies and forces of various structures with DFT calculations, as shown in Fig.~\ref{fig_NN}.


\vspace{12pt}
\begin{center}
\textbf{\large \III\ Soft phonon modes and their symmetry characters}
\end{center}

\begin{figure}[htbp]
	\centering
	\includegraphics[width=3.5in]{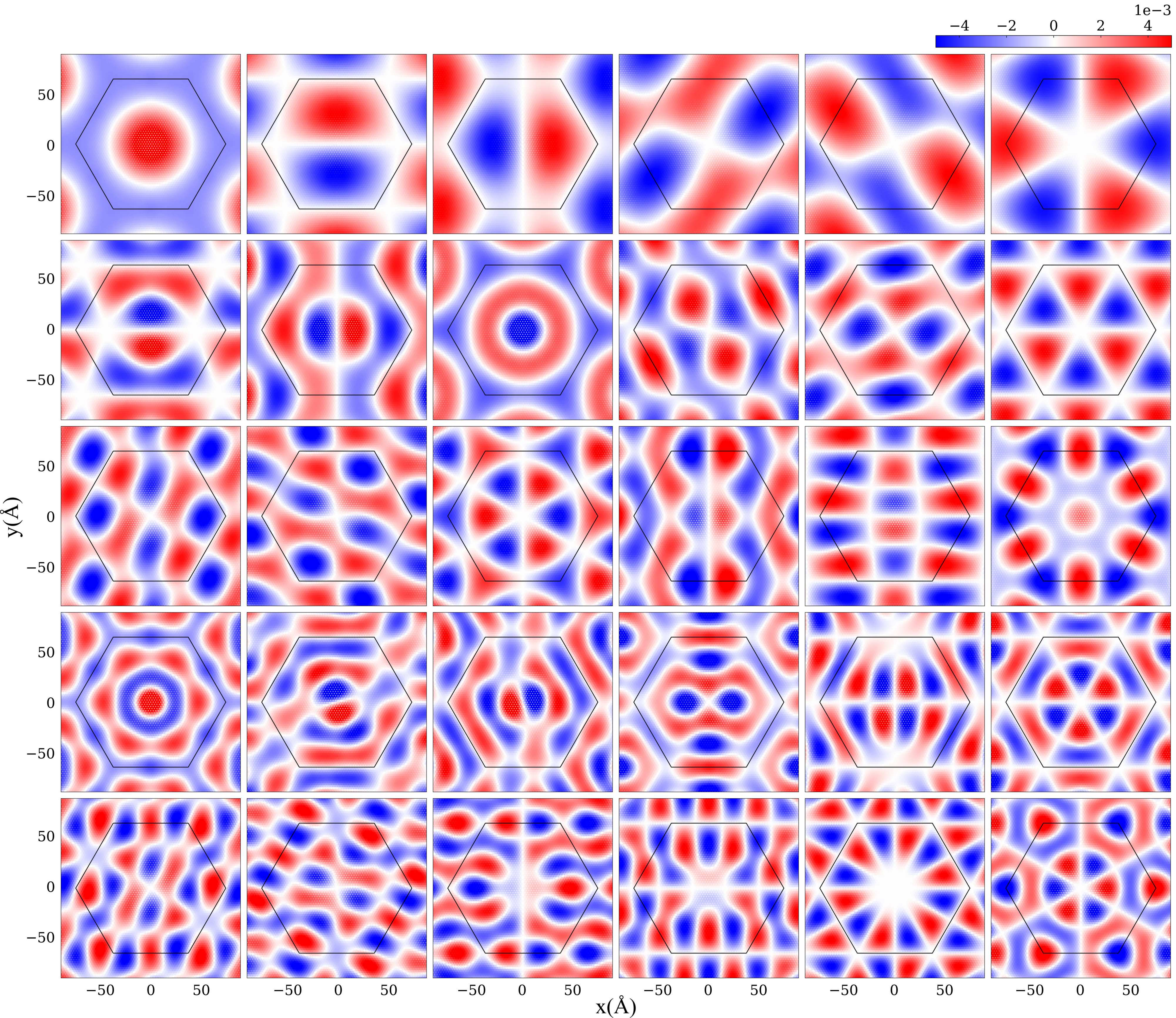}
	\caption{The first 30 soft phonon modes at $\Gamma$} \label{fig_Gamma}
\end{figure}

\begin{figure}[htbp]
	\centering
	\includegraphics[width=3.5in]{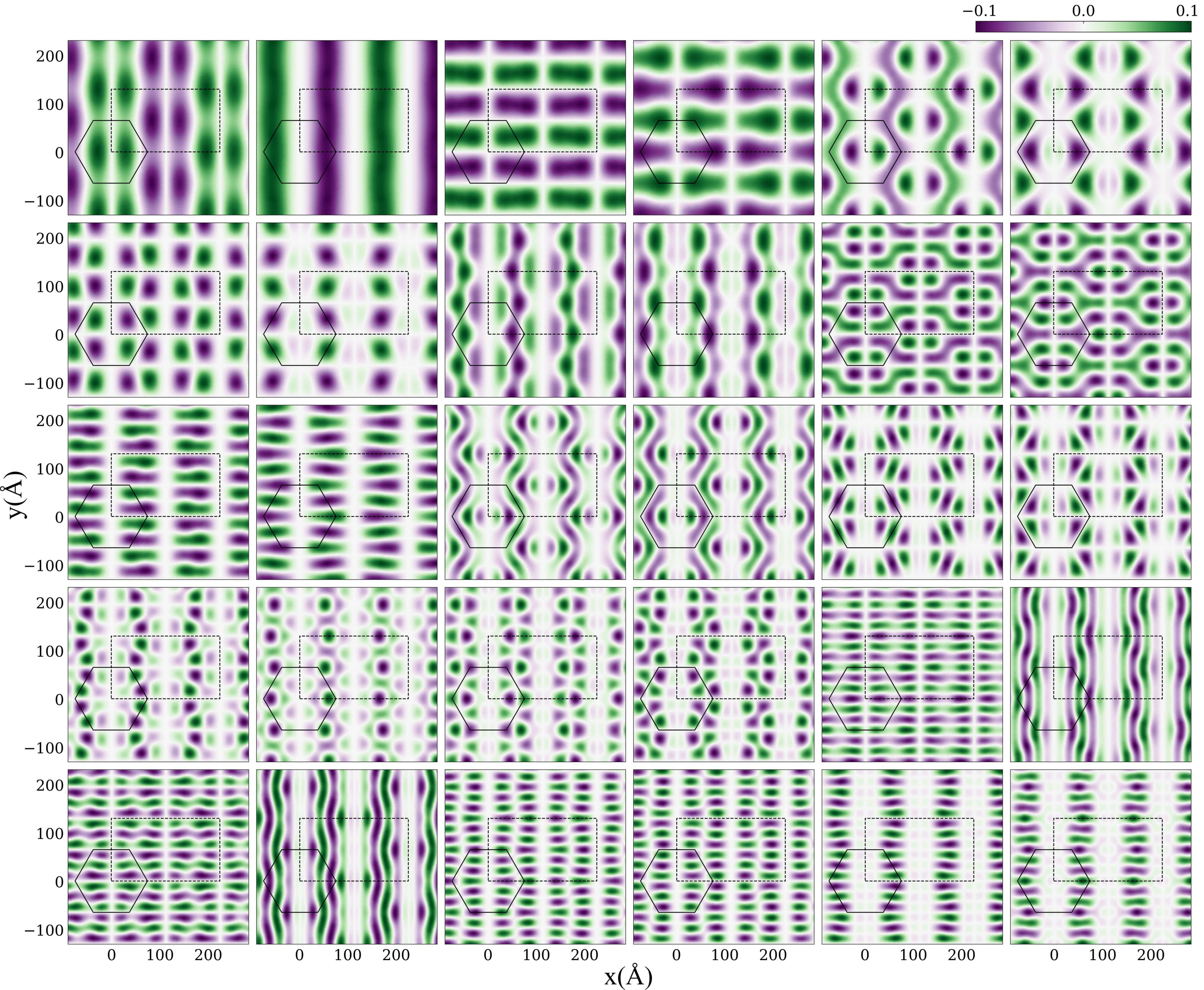}
	\caption{The first 30 soft phonon modes at $M$} \label{fig_M}
\end{figure}

The real-space out-of-plane vibrational patterns of the first 30 soft optical phonon modes of magic-angle TBG at the high symmetry points $\Gamma$ and $M$ have been shown in Fig.~\ref{fig_Gamma} and Fig.~\ref{fig_M} respectively.  We further present the irreducible representations of the first 300 optical phonon modes at $\Gamma$, $M$, and $K$ in \ref{tab:table1}.

\begin{table*}
	\caption{\label{tab:table1}The first 300 phonon modes in $\Gamma$, $M$, and $K$. Such representation is the point group in Schoenflies notation. The first 30 phonon modes have been shown in here, and the whole table could be seen in https://github.com/Simon-lxq/Phonon-symmetry}
		\begin{tabular}{p{3cm}|p{3cm}|p{3cm}|p{3cm}|}
			eigen No.&$\Gamma$&M&K\\ \hline
			1&$D_6$&$C_2$&$C_6$\\
			2&$D_6$&$C_2$&$D_6$\\
			3&$D_6$&$D_2$&$D_6$\\
			4&$C_6$&$C_2$&$C_2$\\
			5&$C_2$&$C_2$&$D_6$\\
			6&$C_2$&$C_2$&$C_6$\\
			7&$D_2$&$C_2$&$D_6$\\
			8&$D_2$&$D_2$&$C_2$\\
			9&$D_3$&$C_2$&$C_6$\\
			10&$C_2$&$C_2$&$D_2$\\
			11&$C_2$&$D_2$&$C_2$\\
			12&$C_6$&$C_2$&$D_3$\\
			13&$D_2$&$C_2$&$C_6$\\
			14&$D_2$&$C_2$&$C_2$\\
			15&$D_3$&$C_2$&$C_2$\\
			16&$D_2$&$C_2$&$C_2$\\
			17&$D_2$&$D_2$&$C_1$\\
			18&$D_3$&$C_2$&$D_6$\\
			19&$C_2$&$C_2$&$C_6$\\
			20&$C_2$&$C_2$&$D_6$\\
			21&$C_6$&$C_2$&$D_6$\\
			22&$C_6$&$D_2$&$D_6$\\
			23&$C_2$&$C_2$&$D_3$\\
			24&$C_2$&$C_2$&$C_2$\\
			25&$D_2$&$D_2$&$D_3$\\
			26&$D_2$&$C_2$&$C_1$\\
			27&$D_3$&$C_2$&$C_6$\\
			28&$D_2$&$C_2$&$C_2$\\
			29&$D_2$&$D_2$&$C_2$\\
			30&$C_2$&$C_2$&$C_6$\\
		\end{tabular}
\end{table*}

\vspace{12pt}
\begin{center}
\textbf{\large \IV\ Band structures under frozen phonon approximation}
\end{center}

\begin{figure}[htbp]
	\centering
	\includegraphics[width=5in]{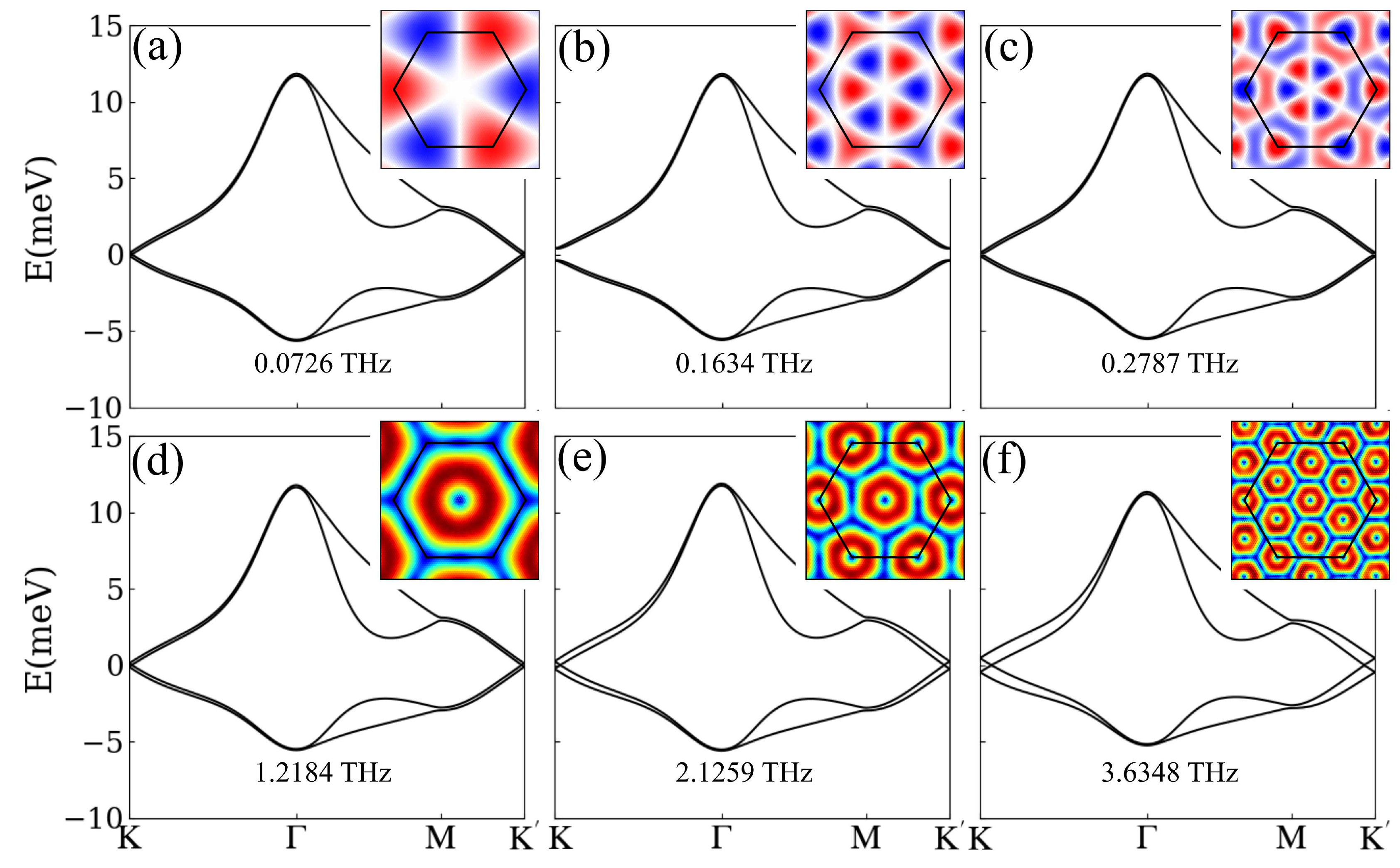}
	\caption{The energy bands of typical phonon modes in high symmetry point $\Gamma$. (a)-(c) The band structure of the soft modes with the frequencies $0.0726\,$THz,$0.1634\,$THz,$0.2387\,$THz, respectively. The insets are the general patterns of out-of-plane vibrational amplitudes, where the average amplitudes is 0.05\,\angstrom\. (d)-(f) The band structure of the soft modes with the frequencies $1.2184\,$THz,$2.1259\,$THz,$3.6348\,$THz, respectively. The insets are the general patterns of in-plane vibrational amplitudes.} \label{fig_band}
\end{figure}

The energy bands with some typical soft phonon modes being brozen are shown in Fig.~\ref{fig_band}. In particular, in Fig.~\ref{fig_band}(a)-(c) we show the band structures with out-of-plane ``octupolar" phonons assumed to be frozen, while in Fig.~\ref{fig_band}(d)-(f) we show the band structures with some ``vortical" phonon modes being frozen (the phonon modes are shown in the insets).  Our calculations indicate that the octupolar phonon modes tend to open up a gap at the charge neutraility point due to the broken $C_{2z}$ symmetry, while the ``vortical" phonon modes tend to enhance intervalley scattering, which splits two Dirac points.

\vspace{12pt}
\begin{center}
\textbf{\large \V\ Lattice vectors and reciprocal lattice vectors of a commensurate supercell}
\end{center}

\begin{figure}[htbp]
	\centering
	\includegraphics[width=5in]{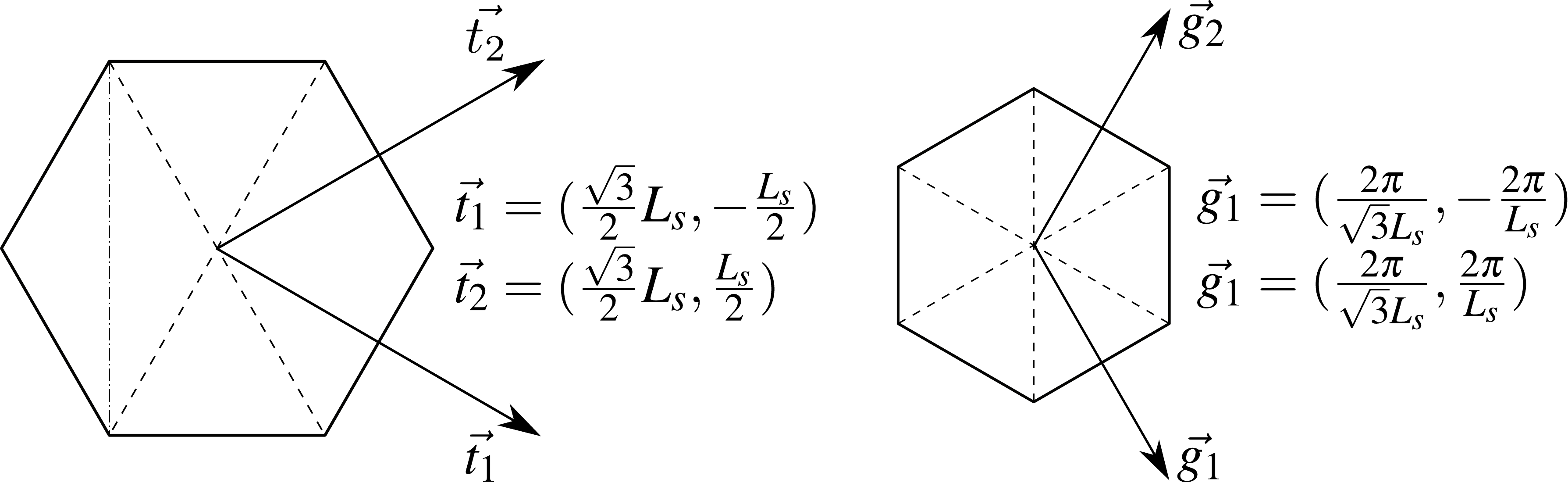}
	\caption{Real space base vector and inverted space base vector} \label{fig_MBZ}
\end{figure}

We consider studying a phonon mode at moir\'e wavevector $\mathbf{q}$, which can be written as
\begin{equation}
	\mathbf{q} = \frac{q_1}{p_1}\mathbf{g_1}+\frac{q_2}{p_2}\mathbf{g_2}\label{e1}
\end{equation}
where $(q_1,p_1)$ and $(q_2,p_2)$ are two pairs of co-prime integers. 
In order to plot the real-space vibrational pattern of the phonon mode at wavevector $\mathbf{q}$, we need to construct a minimal real-space supercell with the superlattice vectors:
\begin{equation}
\begin{split}
	\mathbf{R_1} = n_{11}\mathbf{t_1}+n_{12}\mathbf{t_2} \\ 
	\mathbf{R_2} = n_{21}\mathbf{t_1}+n_{22}\mathbf{t_2}\label{e2}
\end{split}
\end{equation}
We require $\mathbf{R_1}\cdot\mathbf{q}=2\pi, \mathbf{R_1}\cdot\mathbf{q}=0$, such that
\begin{equation}
\begin{split}
	\frac{n_{11}q_1}{p_1}+\frac{n_{12}q_2}{p_2}=1\\
	\frac{n_{21}q_1}{p_1}+\frac{n_{22}q_2}{p_2}=0\label{e3}
\end{split}
\end{equation}
And from Eq.\eqref{e3} we obtain:
\begin{equation}
\begin{split}
	\mathbf{G_1}=\frac{n_{22}\mathbf{g_1}-n_{21}\mathbf{g_2}}{n_{11}n_{22}-n_{12}n_{21}}\\
	\mathbf{G_2}=\frac{n_{11}\mathbf{g_2}-n_{12}\mathbf{g_1}}{n_{11}n_{22}-n_{12}n_{21}}\label{e4}
\end{split}
\end{equation}
We find integer solutions of Eq.\eqref{e3}, and construct a supercell with minimal area. The area of the primitive supercell $\Omega_S$ is expressed as:
\begin{equation}
	\Omega_S = (\mathbf{R_1}\times\mathbf{R_2})\cdot\hat{z}=(n_{11}n_{22}-n_{12}n_{21})\Omega_0\label{e5}
\end{equation}

\end{document}